\documentclass[doublecol]{epl2} 

\title{The Tully-Fisher's law and Dark Matter effects derived via modified symmetries}
\shorttitle{Title} 

\author{Ivan Arraut \inst{1}}
\shortauthor{Ivan Arraut}

\institute{                    
  \inst{1}Institute of Science and Environment and FBL,\\
 University of Saint Joseph,\\
 Estrada Marginal da Ilha Verde, 14-17, Macao, China.
}

\abstract{In any physical system, when we move from short to large scales, new spacetime symmetries emerge which help us to simplify the dynamics of the system. In this letter we demonstrate that certain variations on the symmetries of General Relativity at large scales, generate the effects equivalent to Dark Matter. In particular, we reproduce the Tully-Fisher law, consistent with the predictions proposed by MOND. Additionally, we demonstrate that the dark matter effects derived in this way, are consistent with the predictions suggested by MOND, without modifying gravity.}

\begin{document}

\maketitle

\section{Introduction}

General Relativity (GR) marked, together with Quantum Mechanics, a scientific revolution early at the twentieth century \cite{1,2}. By the time of its discovery, the theory solved some important puzzles \cite{1}. Subsequently, important predictions of the theory were proved experimentally \cite{3}. GR was able to explain the precession of the orbit of Mercury, gravitational waves, the dragging effect, the gravitational time-dilation effect, the gravitational lenses, among  many other cosmological and astrophysical effects which have been tested experimentally \cite{3, 4, 5, 6, 7}. Yet still there are certain observations which GR in its standard form, has not been able to explain by itself. Among these observations, we have the effects normally attributed to Dark Matter and those attributed to Dark Energy \cite{8, 9}. If General Relativity is correct, then the Dark Matter effects should come from some kind of matter which is invisible for all the interactions, except gravity \cite{8}. However, alternative theories of gravity has been formulated in order to explain the Dark Matter effects, but all of them present serious problems difficult to solve \cite{10, 12, 13, 14, 15, 16, 1000, 1001}. Additionally, the theory of MOND, although it is able to make important predictions about the dynamics of the galaxies, being in this way an alternative to Dark Matter \cite{11, 111, 112, 113}, by the date it does not have a valid (free of pathologies) relativistic version able to explain the observed gravitational lenses \cite{12, 13, 14, 15, 16}. In this paper, without modifying gravity, we explain how the MONDian effects emerge naturally from the fact that the symmetry under rotations at galactic scales, is not satisfied anymore and instead a new (modified) symmetry emerges. The new symmetry is equivalent to a modification of the Keppler's law \cite{1}, which changes from its version of equal areas in equal times toward equal arcs in equal time. By imposing the new symmetry, the Tully-Fisher law emerges naturally \cite{TF}. Additionally, with the same argument we can explain why the Dark Matter effects emerge when the accelerations of the bodies involved are very small. This is the same argument used by the theory of MOND in order to explain the Dark Matter effects \cite{11, 111, 112, 113}. Additionally, we show that the low-acceleration condition has to be complemented with a large angular momentum condition, to be explained in this paper. Interestingly, the proposed formalism does not modify the theory of GR and it is completely based on symmetry arguments. Finally, we briefly explore the gravitational lenses, explaining from this perspective why does the observed enhancement of the gravitational interaction occur.

\section{Standard General Relativity: Einstein equations}

The standard Einstein equations can be expressed as \cite{3}

\begin{equation}   \label{Einstein1}
R_{\mu\nu}-\frac{1}{2}g_{\mu\nu}R=8\pi G T_{\mu\nu}.  
\end{equation}
Here $R_{\mu\nu}$ is the Ricci tensor, $R=g_{\mu\nu}R^{\mu\nu}$ is the curvature scalar. Additionally, $G$ is the Newtonian constant and $T_{\mu\nu}$ is the energy-momentum tensor (source term). The equation (\ref{Einstein1}) is able to explain the observed universe at solar system scales, the existence of Black-Holes, the existence of Gravitational Waves, among other observations \cite{3, 4, 5, 6, 7}. Without any modification, eq. (\ref{Einstein1}) can also explain the galaxy rotation curves and the observed gravitational lenses if we introduce Dark-Matter as a source inside the energy-momentum tensor $T_{\mu\nu}$ \cite{8}. This requires the assumption of the existence of matter that cannot be seen and which cannot feel any interaction, except the gravitational interaction \cite{8}. Eq. (\ref{Einstein1}) can be obtained from the Einstein-Hilbert (EH) action expressed as follows

\begin{equation}   \label{Action}
S=\frac{1}{2\kappa}\int d^4x\sqrt{-g}R+S_M.
\end{equation}
Here $S_M$ is the matter action related to the source term $T_{\mu\nu}$ in eq. (\ref{Einstein1}). In addition, $\kappa=8\pi G$  and $g$ is the determinant of the metric $g_{\mu\nu}$ \cite{1}. 

\subsection{Symmetries of the spherically symmetric metric}

Here we consider the standard spherically symmetric metric, namely, the Schwarzschild metric defined as follows

\begin{equation}   \label{m1}
ds^2=-\left(1-\frac{2GM}{r}\right)dt^2+\left(1-\frac{2GM}{r}\right)^{-1}dt^2+r^2d\Omega^2.    
\end{equation}
It is a trivial task to derive the geodesics based on this metric. The geodesics are simplified after considering the symmetries derived from the metric \cite{1}. The symmetries here emerge from the definitions of Killing vectors. The Killing vectors define conserved quantities through the relation \cite{1}

\begin{equation}   \label{Kequa}
K_\mu\frac{dx^\mu}{d\lambda}=C,    
\end{equation}
where $C$ is a constant of motion, $K_\mu$ is the one-form emerging from the Killing vector $K^\nu$ and $\frac{dx^\mu}{d\lambda}$ is the standard derivative of the spacetime coordinates with respect to the affine parameter $\lambda$. For the spherically symmetric case, we have four constants of motion; two of them are related to the direction of the angular momentum, another conserved quantity is related to the magnitude of the same angular momentum and finally, there is a conserved quantity which corresponds to the energy conservation. Once we fix the plane of rotation of the bodies under analysis, then we only have to worry about the magnitude of the angular momentum and on the conserved quantity related to the energy conservation. It is a trivial task to demonstrate that the conserved energy is given by 

\begin{equation}   \label{E}
E=\left(1-\frac{2GM}{r}\right)\frac{dt}{d\lambda},
\end{equation}
and the conserved quantity related to the magnitude of the angular momentum is

\begin{equation}   \label{phi}
L=r^2\frac{d\phi}{d\lambda}.    
\end{equation}
This equation can be interpreted as the Keppler's law, which suggests that when a test body rotates around the center source in a system, then it covers equal areas at equal times \cite{1}. The equations of motion for this system are obtained from the expansion $g_{\mu\nu}dx^\mu dx^\nu=-1$ (for massive particles), complemented with the conserved quantities defined in eqns. (\ref{E}) and (\ref{phi}). The final result is \cite{1}

\begin{equation}   \label{motion}
\frac{1}{2}\left(\frac{dr}{d\lambda}\right)^2+V(r)=\Gamma,    
\end{equation}%
with $\Gamma=\frac{1}{2}E^2-\frac{1}{2}$. In eq. (\ref{motion}), the potential $V(r)$ is defined as 

\begin{equation}   \label{Potential}
V(r)=-\frac{GM}{r}+\frac{L^2}{2r^2}-\frac{GML^2}{r^3}.    
\end{equation}
The first term of this potential corresponds to the Newtonian contribution, the second corresponds to the dynamical effects due to the centrifugal contribution and finally, the last term emerges exclusively from GR (it does not appear in Newtonian gravity) and it is the term coupling the source term $GM$ with the angular momentum $L$. This term is the one explaining the observed precession of perihelion of Mercury. It is important to remark that the potential in eq. (\ref{Potential}) applies for massive test particles and for massless particles the first term, namely, $-\frac{GM}{r}$ is not considered. 

\section{Modifications of the Keppler's law at large scales}

Although it is well-known that our spacetime is four-dimensional, when we go from short scales toward large scales, the spacetime symmetries might change and these changes modify the dynamic of the system. Then for example, at galactic scales, the angular momentum is not the conserved quantity to consider when we analyze the geodesics. Instead, the conserved quantity replacing eq. (\ref{phi}) is

\begin{equation}   \label{Newconserved}
\frac{L^2}{r}=\gamma^2. 
\end{equation}
This is a modification of the Kepler's law at galactic scales. The result (\ref{Newconserved}) is equivalent to suggest that the conserved quantity at galactic scales is not the angular momentum but rather the velocity, which is what has been perceived at galactic scales in agreement with the observations \cite{8, 11, 111, 112, 113}. A way to visualize this aspect is to analyze the Killing vector related to the symmetries under spatial rotations (angular momentum conservation). Since the velocity is the new conserved quantity, then the expression (\ref{phi}) is replaced by the new conserved quantity

\begin{equation}   \label{Newconserved2}
\gamma=r\left(\frac{d\phi}{d\lambda}\right).
\end{equation}
If we use the result (\ref{phi}) on the previous equation, we then obtain eq. (\ref{Newconserved}). Since the conservation of the angular momentum comes out from the expression defined in eq. (\ref{Kequa}), after taking into account that the spatial Killing vector is $K^\mu=(0, 0, 0, 1)$ \cite{1}. The one-form related to this vector is obtained after downloading the index $\mu$ by using the metric as follows

\begin{equation}   \label{Impor}
K_\mu=g_{\mu\nu}K^\nu=r,    
\end{equation}
at galactic scales. Applying then the expression (\ref{Kequa}), we then get the conserved quantity (\ref{Newconserved2}). It is important to remark that the result (\ref{Impor}), can be only obtained if the metric at galactic scales suffers a modification on the angular part, such that $r^2d\Omega^2\to rd\Omega^2$. This is the necessary condition for the velocity to become a conserved quantity (instead of the angular momentum). If we expand the metric with this modification, following the same steps as those giving the results (\ref{motion}) and (\ref{Potential}), then we find out that the dynamics of the galaxy follows the following expression

\begin{equation}   \label{motion2}
\frac{1}{2}\left(\frac{dr}{d\lambda}\right)^2+V_1(r)=\Gamma,    
\end{equation}
with the new potential $V_1(r)$, defined as

\begin{equation}   \label{Potential1}
V_1(r)=-\frac{GM}{r}+\frac{\gamma^2}{2r}-\frac{GM\gamma^2}{r^2}.    
\end{equation}
The zero-condition for the gradient of this equation gives us the the equilibrium condition after using the result $\nabla V_1(r)=0$. In this way, we obtain 

\begin{equation}
-\frac{GM}{r^2}+\frac{\gamma^2}{2r^2}-\frac{2GM\gamma^2}{r^3}=0.
\end{equation}
The solution for this equation is

\begin{equation}   \label{Diverg}
r_{eq}=\frac{4GM\gamma^2}{\gamma^2-2GM}.    
\end{equation}
In a moment we will explain why the dark matter effects emerge when $\gamma^2\to2GM$, which gives an apparent divergence in eq. (\ref{Diverg}). For understanding more about this apparent divergence and the related scales emerging from the relation, we have to consider that $\gamma^2=L^2/r$. Replacing this condition over eq. (\ref{Diverg}), gives the following quadratic equation

\begin{equation}
r_{eq}^2-\frac{L^2}{2GM}r_{eq}+2L^2=0. 
\end{equation}
Solving this equation gives us the solution

\begin{equation}
r_{eq}=\frac{L^2}{4GM}\left(1\pm\sqrt{1-32\left(\frac{GM}{L}\right)^2}\right).    
\end{equation}
This solution gives us a minimal value for the angular momentum of $L_{min}=4\sqrt{2}GM$ with the corresponding equilibrium radius of $r_{eq}=8GM$. However, this regime is not the interesting one for the purposes of this analysis. The things come out to be more interesting when we consider the regime of large angular momentum defined by the condition $L>>GM$. In this case, we have two solutions for $r_{eq}$. The first one is $r_{eq1}\approx4GM$ and the second one is

\begin{equation}   \label{secsol}
r_{eq2}\approx\frac{L^2}{2GM}.    
\end{equation}
This means that $L^2/r_{eq}\approx2GM$, which is precisely the condition $\gamma^2\to2GM$ after considering the result (\ref{Newconserved}). 

Taking into account that $L=rv$ (taking unitary value for the mass of the test particle), we then get the general form of the Tully-Fisher's law, here given as 

\begin{equation}   \label{MOND1}
r_{eq2}=\frac{2GM}{v^2}.    
\end{equation}
The MOND regime appears when the acceleration $\frac{v^2}{r}\to a_0$, where $a_0$ is some pre-determined scale. By using this acceleration scale inside eq. (\ref{MOND1}), we get

\begin{equation}
r_{eq2}\approx\sqrt{\frac{GM}{a_0}},   
\end{equation}
which is the well-known scale at which the MONDian regime operates. If we use this result in eq. (\ref{MOND1}), we get

\begin{equation}
v^4=4GMa_0.    
\end{equation}
This is a more explicit form of the Tully-Fisher's law in the MONDian language \cite{10, 11, 111, 112, 113}. It is interesting to notice that the key term in these calculations is the term coupling the angular momentum $L$ with the source term $GM$ inside the potential (\ref{Potential1}). This term does not appear in Newtonian gravity \cite{1}. We must remark that the Dark Matter effects appear at low accelerations $\frac{v^2}{r}\to a_0$, but with the simultaneous condition of large angular momentum ($L>>GM$), which brings out the solution (\ref{secsol}). Finally, we would like to remark here the gravitational enhancement obtained at the MONDian regime. The terms generating gravitational attraction in eq. (\ref{Potential1}) can be combined as

\begin{equation}   \label{prevequ}
V_{1Att}(r)=-\frac{GM}{r}\left(1+\frac{\gamma^2}{r}\right)=-\frac{GM_{eff}}{r},    
\end{equation}
after considering the acceleration limit $a_0$. In this previous equation, the subindex $Att$ means "Attractive" and we have also defined $M_{eff}=1+\frac{\gamma^2}{r}$, which can be interpreted as an effective mass at the moment of calculating gravitational lenses. This effective mass is enhanced by the term coupling the angular momentum with the source term in eq. (\ref{Potential1}) and this could explain the observed enhancements via gravitational lenses. Then it is evident that the deflection angle of the light crossing a galaxy will be enhanced by the third term of the potential in eq. (\ref{Potential1}).

\section{Conclusions}

In this paper we have demonstrated that the Dark Matter effects emerge naturally from the standard theory of GR, but considering a modification of the conserved quantity associated to spatial rotations at large scales. Interestingly, not only the Tully-Fisher's law and the MONDian effects emerge naturally but also the gravitational enhancements, necessary for reproducing additional deflection angles when we consider gravitational lenses, appear naturally from the same formulation. We have also proved that the term coupling the angular momentum with the source term is the main responsible for the emergence of the Dark Matter effects, after considering the modified symmetry under spatial rotations. The modification of this symmetry suggests that the Keppler's law does not follow the standard format at galactic scales. Then the test bodies instead of sweeping equal areas at equal times, they sweep equal arcs at equal times at galactic scales. This modification is enough for reproducing the known Dark Matter effects without modifying gravity. Finally, it is important to remark that we have demonstrated that for getting the Dark Matter effects, not only low accelerations are necessary, but in addition large magnitudes of the angular momentum for the bodies moving around the center of the galaxy. These deep details are not fully explained inside the standard MONDian formulation, which is fully based on the acceleration regime and an unknown interpolating function. The interpolating function in such a case, turns our to make corrections to the Newtonian gravity at galactic scales due to the low accelerations of the objects at those scales \cite{11, 111, 112, 113}.



\begin{thebibliography}{0}

\bibitem{1}
Carrol, S.; {\it Spacetime and Geometry: An Introduction to General Relativity}, San Francisco: Addison-Wesley, ISBN 978-0-8053-8732-2.

\bibitem{2}
Zettili, N.; {\it Quantum Mechanics: Concepts and Applications}, Wiley; 2nd edition (February 24, 2009), ISBN-10:0470026790; ISBN-13:978-0470026793.

\bibitem{3}
Einstein, A.; {\it The Foundation of the General Theory of Relativity}, Annalen der Physik. 49 (7): 769–822.

\bibitem{4}
Castelvecchi, D. and Witze, A.; {\it Einstein's gravitational waves found at last}, Nature (2016).   

\bibitem{5}   
Abbot, B. P.; et al. (LIGO Scientific Collaboration and Virgo Collaboration) (2016), {\it Observation of Gravitational Waves from a Binary Black Hole Merger}, Phys. Rev. Lett. {\bf 116} (6): 061102.

\bibitem{6}
Ciufolini and Pavlis, E. C. (2004), {\it A confirmation of the general relativistic prediction of the Lense–Thirring effect}, Nature. {\bf 431} (7011): 958–960.

\bibitem{7}
Will, C.M. (2006), {\it The Confrontation between General Relativity and Experiment}, Liv. Rev. in Rel. {\bf 9} (1): 39.

\bibitem{8}
Trimble, V. (1987), {\it Existence and nature of dark matter in the universe}, Annual Review of Astronomy and Astrophysics {\bf 25}: 425–472.

\bibitem{9}
Peebles, P. J. E. and Ratra, Bharat (2003), {\it The cosmological constant and dark energy}, Rev. of Mod. Phys. {\bf 75} (2): 559–606.
  
\bibitem{10}
Arraut, I.; {\it Can a nonlocal model of gravity reproduce Dark Matter effects in agreement with MOND?}, Int.J.Mod.Phys.D {\bf 23} (2014) 1450008.

\bibitem{12}
Bekenstein, J. D. and Sanders, R. H.; {\it TeVeS/MOND is in harmony with gravitational redshifts in galaxy clusters}, 	Mon. Not. Roy. Astron. Soc. {\bf 421}, L59-L61 (2012).

\bibitem{13}
Bekenstein, J. D. (2004), {\it Relativistic gravitation theory for the modified Newtonian dynamics paradigm}, Physical Review D, 70 (8): 083509.

\bibitem{14}
Bekenstein, J. D. and Milgrom, M.; {\it Does the missing mass problem signal the breakdown of Newtonian gravity?}, Astrophys. Journ. {\bf 286}, 7 (1984). 

\bibitem{15}
Bekenstein, J. D.; {\it Relativistic gravitation theory for the MOND paradigm}, Nucl.Phys. {\bf A827} (2009) 555C. 

\bibitem{16}
Bekenstein, J. D.; {\it Relativistic MOND as an alternative to the dark matter paradigm}, Nucl. Phys. A. {\bf 827}: 555c–560c, (2009).

\bibitem{1000}
J. Khoury, {\it Alternative to particle dark matter},
Phys. Rev. D, {\bf91} (2015).

\bibitem{1001}
L. Berezhiani, J. Khoury, {\it Theory of dark matter superfluidity}, Phys. Rev. D, {\bf 92} (2015), p. 103510.

\bibitem{11}
Milgrom, M.; {\it A Modification of the Newtonian Dynamics: Implications for Galaxy systems}, Astrophys.{\bf J. 270} (1983) 384

\bibitem{111}
Milgrom, M.; {\it A Modification of the Newtonian dynamics: Implications for galaxies}, Astrophys. {\bf J. 270} (1983) 371.

\bibitem{112}
Milgrom, M.; {\it A Modification of the Newtonian dynamics as a possible alternative to the hidden mass hypothesis}, Astrophys. {\bf J. 270} (1983) 365. 

\bibitem{113}
Milgrom, M.; {\it A Modification of the Newtonian dynamics as a possible alternative to the hidden mass hypothesis}, Astrophys. {\bf J 270} (1983), 365.

\bibitem{TF}
Tully, R. B. and Fisher, J. R. {\it A New Method of Determining Distances to Galaxies}, Astron. and Astrophys. {\bf 54} (3): 661–673, (1977).

\end{thebibliography}
\end{document}